\documentclass[12pt]{article}
\usepackage{amssymb,amsmath,epsfig}

\begin{document}

\title{\bf Shearfree Condition and dynamical Instability in $f(R,T)$ gravity}

\author{Ifra Noureen \thanks{ifra.noureen@gmail.com} ${}^{(a)}$,  M. Zubair
\thanks{drmzubair@ciitlahore.edu.pk} ${}^{(b)}$,  A.A. Bhatti
\thanks{drabhatti@umt.edu.pk} ${}^{(a)}$, G. Abbas \thanks{gulamabbas@ciitsahiwal.edu.pk} ${}^{(c)}$\\
${}^{(a)}$ Department of Mathematics, SST,\\University of Management and
Technology, Lahore-54770, Pakistan.\\ ${}^{(b)}$ Department of Mathematics,\\
COMSATS Institute of Information Technology, Lahore-54700, Pakistan.\\
${}^{(c)}$ Department of Mathematics,\\ COMSATS
Institute of Information Technology, Sahiwal-57000, Pakistan}
\date{}
\maketitle

\begin{abstract}
The implications of shearfree condition on instability range of anisotropic
fluid in $f(R,T)$ are studied in this manuscript. A viable $f(R, T)$ model is
chosen to arrive at stability criterion, where $R$ is Ricci scalar and $T$ is
the trace of energy momentum tensor. The evolution of spherical star is
explored by employing perturbation scheme on modified field equations and
contracted Bianchi identities in $f(R, T)$. The effect of imposed shearfree
condition on collapse equation and adiabatic index $\Gamma$ is studied in
Newtonian and post-Newtonian regimes.
\end{abstract}


\section{Introduction}

In a recent work \cite{1}, we have studied the effect of anisotropic fluid on
dynamical instability of spherically symmetric collapsing star in $f(R, T)$
theory. Herein, we plan to explore the instability range of anisotropic
spherically symmetric stars, considering shearfree condition. The role of
shear tensor in evolution of gravitating objects and consequences of
shearfree condition has been studied extensively. Collins and Wainwright
\cite{2} studied the impact of shear on general relativistic cosmological and
stellar models. Herrera et al. \cite{3,4} worked out the homology and
shearfree conditions for dissipative and radiative gravitational evolution.

The features of gravitational evolution and its final outcome is of
great importance in view of General Relativity (GR) as well as in
modified theories of gravity. Shearfree collapse accounting heat
flow is discussed in \cite{5}, it is established that shear plays
critical role in gravitational evolution and may lead to the
formation of naked singularities \cite{6}. It is mentioned in
\cite{6} that occurrence of shearing effects near collapsing stars
detains the apparent horizon leading to formation of naked
singularity. However vanishing shear give rise to the formation of
apparent horizon and so the evolving cloud ends in a black hole
(BH). Thus, relevance of shear tensor in structure formation and its
impressions on dynamical instability range of self gravitating body
is well motivated direction of study.

Stars shine by consuming their nuclear fuel, continuous fuel consumption
cause imbalance between inward acting gravitational pull and outward drawn
pressure giving rise to collapse \cite{7}. The outcome of gravitational
evolution is size dependent as well as other physical aspects \cite{8} such
as isotopy, anisotropy, shear, radiation, dissipation etc. In comparison to
the stars of mass around one solar masses, massive stars tend to lose nuclear
fuel more rapidly and so more unstable. Pressure to density ratio naming
adiabatic index, denoted by $\Gamma$ is utile in estimation of
stability/instability range of stars. Chandrasekhar \cite{9} explored the
instability range of spherical stars in terms of $\Gamma$.

Herrera and his contemporaries \cite{10}-\cite{17} contributed majorally in
addressing the instability problem in general relativity (GR), accompanying
various situations, i.e., isotopy, anisotropy, shearfree condition,
radiation, dissipation, expansionfree condition, shearing expansionfree
fluids. In order to achieve the more precise and generic description of
universe, the dark energy components are incorporated by introducing modified
theories of gravity. Modified theories are significant in advancements
towards accelerated expansion of the universe and to present corrections to
GR on large scales. The modifications are introduced in Einstein Hilbert (EH)
action by inducing minimal or non-minimal coupling of matter and geometry
\cite{17a}.

Dynamical analysis of self gravitating sources in modified theories of
gravity has been discussed extensively in recent years. The null dust
non-static exact solutions in $f(R)$ gravity are studied in \cite{18},
Cembranos et al. \cite{19} studied the evolution of gravitating sources in
the presence of dust fluid. Instability range of spherically and axially
symmetric anisotropic stars has been established in context of $f(R)$ gravity
\cite{20, 21}, concluding that deviations from spherical  symmetry
complicates the subsequent evolution.

Harko et al. \cite{22} presented the $f(R,T)$ theory of gravity as another
alternate to GR and generalization of $f(R)$ theory representing non-minimal
matter to geometry coupling. The action in $f(R,T)$ gravity includes
arbitrary function of Ricci scalar $R$ and trace of energy momentum tensor
$T$ to take into account the exotic matter. After the introduction of
$f(R,T)$ gravity, its cosmological and thermodynamic implications were widely
studied \cite{23}-\cite{26} including the energy conditions. Recently, we
have studied the evolution of anisotropic gravitating source with zero
expansion \cite{27}. Herein, we are interested in exploration of shearfree
condition implications on spherically symmetric gravitating source in
$f(R,T)$ gravity.

The modified action in $f(R,T)$ is as follows \cite{22}
\begin{equation}\label{1}
\int dx^4\sqrt{-g}[\frac{f(R, T)}{16\pi G}+\mathcal{L} _ {(m)}],
\end{equation}
where $\mathcal{L} _ {(m)}$ denote matter Lagrangian, and $g$ represents the
metric tensor. Various choices of $\mathcal{L} _ {(m)}$ can be taken into
account, each of which leads to a specific form of fluid. Many people worked
out this problem in GR and modified theories of gravity, stability of general
relativistic dissipative axially symmetric and spherically symmetric with
shearfree condition has been established in \cite{28, 29}. Dynamical analysis
of shearfree spherically symmetric sources in $f(R)$ gravity is presented in
\cite{30}.

The manuscript arrangement is: Section \textbf{2} comprises of modified
dynamical equations in $f(R,T)$ gravity. Section \textbf{3} includes model
under consideration, perturbation scheme and corresponding collapse equation
alongwith shear-free condition in Newtonian and post Newtonian eras. Section
\textbf{4} contains concluding remarks followed by an appendix.

\section{Dynamical Equations in $f(R,T)$}

In order to study the implications of shearfree condition on evolution of
spherically symmetric anisotropic sources, modified field equations in
$f(R,T)$ gravity are formulated by varying action (\ref{1}) with the metric
$g_{uv}$. Here, we have taken $\mathcal{L} _{(m)}= \rho$ \cite{r5}, for this
choice of $\mathcal{L}_{(m)}$ modified field equations in $f(R,T)$ gravity
takes following form
\begin{eqnarray}\nonumber
G_{uv}&=&\frac{1}{f_R}\left[(f_T+1)T^{(m)}_{uv}-\rho g_{uv}f_T+\frac{f-Rf_R}{2}g_{uv}
\right.\\\label{4}&+&\left.(\nabla_u\nabla_v-g_{uv}\Box)f_R\right].
\end{eqnarray}
Here $T^{(m)}_{uv}$ is energy momentum tensor for usual matter taken to be locally
anisotropic.

The three dimensional spherical boundary surface $\Sigma$ is considered that
constitutes two regions named as interior and exterior spacetimes. The line
element for region inside the boundary $\Sigma$ is
\begin{equation}\label{1'}
ds^2_-=A^2(t,r)dt^{2}-B^2(t,r)dr^{2}-C^2(t,r)(d\theta^{2}+\sin^{2}\theta d\phi^{2}).
\end{equation}
The line
element for region beyond $\Sigma$ is \cite{1}
\begin{equation}\label{25}
ds^2_+=\left(1-\frac{2M}{r}\right)d\nu^2+2drd\nu-r^2(d\theta^2+\sin^{2}\theta
d\phi^{2}),
\end{equation}
where $\nu$ is retarded time and $M$ denote the total mass.

The expression for anisotropic energy momentum tensor $T^{(m)}_{uv}$ is given by
\begin{equation}\label{5}
T^{(m)}_{uv}=(\rho+p_{\bot})V_{u}V_{v}-p_{\bot}g_{uv}+(p_{r}-p_{\bot})\chi_{u}\chi_{v},
\end{equation}
where $\rho$ is energy density, $V_{u}$ describes four-velocity of the fluid,
$\chi_u$ is radial four vector, $p_r$ and $p_{\bot}$ represent the radial and
tangential pressure respectively. These physical quantities are linked as
\begin{equation}\label{3}
V^{u}=A^{-1}\delta^{u}_{0},\quad
V^{u}V_{u}=1,\quad \chi^{u}=B^{-1}\delta^u_1,\quad
\chi^{u}\chi_{u}=-1.
\end{equation}
The shear tensor denoted by $\sigma_{uv}$ is defined as
\begin{equation}\label{d}
\sigma_{uv}=V_{(u;v)}-a_{(u}V_{v)}-\frac{1}{3}\Theta(g_{uv}-V_{u}V_{v}),
\end{equation}
where $a_{u}$ is four acceleration and $\Theta$ is expansion scalar, given by
\begin{equation}\label{dd}
a_{u}=V_{(u;v)}V^{v}, \quad \Theta=V_{;u}^{u}.
\end{equation}
Components of shear tensor are found by variation of Eq. (\ref{d}) and these
are used to find expression for shear scalar in the following form
\begin{equation}\label{ddd}
\sigma=\frac{1}{A}\left(\frac{\dot{B}}{B}-\frac{\dot{C}}{C}\right),
\end{equation}
where dot and prime indicate time and radial derivatives respectively. From shearfree condition we arrive at vanishing shear scalar, i.e., $\sigma=0$,
implying $\frac{\dot{B}}{B}=\frac{\dot{C}}{C}$.

It is worth mentioning here that the expansion scalar and a scalar function described in terms of Weyl tensor and the anisotropy of pressure controls the departure 
from shearfree condition. Such function is related to the Tolman mass and appears in a natural way in the orthogonal splitting of the
Reimann tensor \cite{31}. It is obvious that pressure anisotropy and density inhomogeneities have extensive implications on stability of the shear–free condition 
but it is not intuitively clear that their specific combination affects stability \cite{29}. Generically shearfree condition remains unstable against the presence of 
pressure anisotropy. Alternatively, one can consider such a case that pressure anisotropy and density inhomogeneity are present in a way that the scalar function 
appearing in orthogonal splitting of Reimann tensor vanishes, implying non-homogeneous anisotropic stable shear free flow. Since we are dealing with
fluid evolving under shearfree condition, so we shall make use of this
condition while evaluating  the components of field equations and also in
conservation equations.

The components of modified Einstein tensor are
\begin{eqnarray}\label{f1}
G_{00}&=&\frac{1}{f_R}\left[\rho+ \frac{f-Rf_R}{2}+
\frac{f_R''}{B^2}-\frac{3\dot f_R}{A^2}\frac{\dot{B}}{B}
-\frac{f_R'}{B^2}\left(\frac{B'}{B}-\frac{2C'}{C}\right)\right]
,\\\label{f2}
G_{01}&=&\frac{1}{f_R}\left[\dot{f_R}'
-\frac{A'}{A}\dot{f_R}-\frac{\dot{B}}{B}f_R'\right],\\\nonumber
G_{11}&=&\frac{1}{f_R}\left[p_r+\left(\rho+p_r\right)f_T- \frac{f-Rf_R}{2}+
\frac{\ddot{f_R}}{A^2}-\frac{\dot f_R}{A^2}\left(\frac{\dot{A}}{A}-\frac{2\dot{C}}{C}\right)
\right.\\\label{f3}
&&\left.-\frac{f_R'}{B^2}\left(\frac{A'}{A}+\frac{2C'}{C}\right)\right],
\\\nonumber
G_{22}&=&\frac{1}{f_R}\left[p_\perp+\left(\rho + p_\perp\right)f_T-
\frac{f-Rf_R}{2}+
\frac{\ddot{f_R}}{A^2}-\frac{f_R''}{B^2}-\frac{\dot
f_R}{A^2}\left(\frac{\dot{A}}{A} \right.\right.\\\label{f4}
&&\left.\left.-\frac{\dot{2B}}{B}\right)
-\frac{f_R'}{B^2}\left(\frac{A'}{A}-\frac{B'}{B}+\frac{C'}{C}\right)\right].
\end{eqnarray}
The dynamical equations extracted from the conservation laws are vital in the
study of stellar evolution. The conservation of full field equations is
considered to incorporate the non-vanishing divergence terms, Bianchi
identities are
\begin{eqnarray}\label{bb}
&&G^{uv}_{;v}V_{u}=0,\quad
G^{uv}_{;v}\chi_{u}=0,
\end{eqnarray}
on simplification of Eq.(\ref{bb}), we have dynamical equations as follows
\begin{eqnarray}\label{B1} &&\dot{\rho}-\rho\frac{\dot{f_R}}{f_R}+[1+f_T]\left(3\rho+p_r
+2p_\perp\right)\frac{\dot{B}}{B}+ Z_1(r,t)=0,
\\\nonumber &&
(\rho +p_r)f'_T+(1+f_T)\left\{p'_r+\rho \frac{A'}{A}+
p_r\left(\frac{A'}{A}+2\frac{C'}{C}-\frac{f'_R}{f_R}\right)
-2p_\perp\frac{C'}{C}\right\}\\\label{B2}&&+f_T\left(\rho'-\frac{f'_R}{f_R}\right)+Z_2(r,t)=0,
\end{eqnarray}
where $Z_1(r,t)$ and $Z_2(r,t)$ are provided in appendix as Eqs.(\ref{B3})
and (\ref{B4}) respectively. Deviations from equilibrium in conservation
equations with the time transition leads to the stellar evolution,
perturbation approach is devised to the estimate the instability range.

\section{Perturbation Scheme and Shearfree Condition}

We consider a particular $f(R,T)$ model of the form
\begin{eqnarray}\label{m}
&&f(R, T)= R+\alpha R^2+\lambda T,
\end{eqnarray}
where $\alpha$ and $\lambda$ can be any positive constants. Perturbation
approach is utilized to estimate the instability range of spherical star with
shear free condition. This scheme is utile in determination of more generic
analytical constraints on collapse equation, rather to establish dynamical
analysis of special cases numerically. Also, field equations are highly
nonlinear differential equations, in such scenario application of
perturbation lead to beneficial observations.

It is assumed that initially all quantities are independent of time and with
the passage of time perturbed form depends on both time and radial
coordinates. Taking $0<\varepsilon\ll1$, the physical quantities and their
perturbed form can be arranged as
\begin{eqnarray}\label{41} A(t,r)&=&A_0(r)+\varepsilon
D(t)a(r),\\\label{42} B(t,r)&=&B_0(r)+\varepsilon D(t)b(r),\\\label{43}
C(t,r)&=&C_0(r)+\varepsilon D(t)\bar{c}(r),\\\label{44}
\rho(t,r)&=&\rho_0(r)+\varepsilon {\bar{\rho}}(t,r),\\\label{45}
p_r(t,r)&=&p_{r0}(r)+\varepsilon {\bar{p_r}}(t,r),
\\\label{46}
p_\perp(t,r)&=&p_{\perp0}(r)+\varepsilon {\bar{p_\perp}}(t,r), \\\label{47}
m(t,r)&=&m_0(r)+\varepsilon {\bar{m}}(t,r), \\\label{49'}
R(t,r)&=&R_0(r)+\varepsilon D_1(t)e_1(r),\\\label{50'}
T(t,r)&=&T_0(r)+\varepsilon D_2(t)e_2(r),\\\nonumber
f(R, T)&=&[R_0(r)+\alpha R_0^2(r)+\lambda T_0]+\varepsilon(D_1(t)e_1(r)[1\\\label{51'}&+&2\alpha  R_0(r)]+ D_2(t)e_2(r)),\\\label{52'}
f_R&=&1+2\alpha R_0(r)+\varepsilon 2\alpha D_1(t)e_1(r),\\\label{52'}
f_T&=&\lambda.
\end{eqnarray}
Considering Schwarzschild coordinate $C_0=r$ and implementing perturbation
scheme on vanishing shear scalar implies
\begin{equation}\label{dddd}
\frac{b}{B_0}=\frac{\bar{c}}{r}.
\end{equation}
Using Eqs. (\ref{41})-(\ref{52'}) and (\ref{dddd}) in dynamical equations
i.e., Eqs.(\ref{B1}) and (\ref{B2}) leads to following expressions
\begin{eqnarray}\label{B1p}&&
\dot{\bar{\rho}}+\left[\frac{2e\rho_0}{Y}+\lambda_1\frac{\bar{c}}{r}(2\rho_0+p_{r0}+4p_{\perp0})
+YZ_{1p}\right]\dot{D}=0,
\\\nonumber
&&\lambda_1\left\{\bar{p_r}'+\bar{\rho}\frac{A'_0}{A_0}+\bar{p_r}\left(\frac{A'_0}{A_0}+\frac{2}{r}-\frac{2\alpha
R'_0}{Y}\right)-\frac{2\bar{p_\perp}}{r}\right\}+\lambda\bar{\rho'}+2\alpha\ddot{D}\left[\frac{1}{A_0^2}\left(e'
\right.\right.\\\nonumber &&\left.\left.
+2e\frac{B'_0}{B_0}-\frac{\bar{c}}{r}R'_0\right)+B_0^2(Y)\left\{\frac{e}{B_0^2Y}\right\}'\right]
+D\left[\lambda_1[(\frac{a}{A_0})'(\rho_0+p_{r0})\right.\\\nonumber &&\left.-2(p_{r0}+p_{\perp0})(\frac{\bar{c}}{r})']
-\frac{2\alpha}{Y}\left\{\lambda_1\left(p'_{r0}
+\rho_0\frac{A'_0}{A_0}
+p_{r0}\left(\frac{A'_0}{A_0}-\frac{2\alpha R'_0}{Y}+\frac{2}{r}\right)\right)\right\}
\right.\\\label{B2p}&&\left.+\lambda\left(e'+e[\rho'_0-\frac{2\alpha R'_0}{Y}]\right)+YZ_{2p}\right]
=0,
\end{eqnarray}
where $Z_{1p}$ and $Z_{2p}$ are given in appendix, for the sake of simplicity
we put $Y$ in place of $1+2\alpha R_0$ and $\lambda_1=\lambda+1$, assuming
that $D_1=D_2=D$ and $e_1=e_2=e$. Above mentioned perturbed dynamical
equations and perturbed field equations shall be used to arrive at perturbed
physical quantities such as $\bar{\rho}, \bar{p_r}$ and $\bar{p_\perp}$.

The expression for $\bar{\rho}$ can be found from Eq.(\ref{B1p}), as follows.
\begin{equation}\label{B1p'}
\bar{\rho}=-\left[\frac{2e\rho_0}{Y}+\frac{\bar{c}}{r}(3\rho_0+p_{r0}+4p_{\perp0})+YZ_{1p}\right]D.
\end{equation}
The Harrison-Wheeler type
equation of state relate $\bar{\rho}$ and $\bar{p_r}$, given by
\begin{equation}\label{B7}
\bar{p}_r=\Gamma\frac{p_{r0}}{\rho_0+p_{r0}}\bar{\rho}.
\end{equation}
Putting $\bar{\rho}$ from Eq.(\ref{B1p'}) in Eq.(\ref{B7}), we found
\begin{equation}\label{B8}
\bar{p}_r=-\Gamma\frac{p_{r0}}{\rho_0+p_{r0}}\left[\frac{2e\rho_0}{Y}
+\lambda_1\frac{\bar{c}}{r}(3\rho_0+p_{r0}+4p_{\perp0})+YZ_{1p}\right]D.
\end{equation}
Perturbed form of field equation (\ref{f4}) yields expression for $\bar{p}_\perp$, that turns out to be
\begin{eqnarray}\label{B9}
\bar{p}_\perp=\{\frac{Y\bar{c}}{r}-2\alpha e\}\frac{\ddot{D}}{A_0^2}
-\frac{\lambda \bar{\rho}}{\lambda_1}+\left\{\left(p_{\perp0}
-\frac{\lambda}{\lambda_1}\rho_0\right)\frac{2\alpha e}{Y}+\frac{Z_3}{\lambda_1}\right\}D,
\end{eqnarray}
$Z_3$ is effective part of the field equation given in appendix as
Eq.(\ref{Z3}).

Substitution of $\bar{\rho}, \bar{p_r}$ and $\bar{p_\perp}$ from
Eqs.(\ref{B1p'}), (\ref{B8}) and (\ref{B9}) in Eq.(\ref{B2p}) leads to collapse equation as under
\begin{eqnarray}\nonumber
&&\ddot{D}\left[\frac{2\alpha }{A_0^2Y}\left\{e'+2e\frac{B'_0}{B_0}
-\frac{\bar{c}}{r}R'_0\right\}-2\alpha B_0^2\left\{\frac{e}{B_0^2Y}\right\}' +\frac{1}{A_0^2}\{\frac{Y \bar{c}}{r}
-2\alpha e\}\right]\\\nonumber &&+D\left[\frac{1}{Y}\left\{\lambda_1\left((\rho_0+p_{r0})\left(\frac{a}{A_0}\right)'
-2(\rho_0+p_{\perp0})\left(\frac{\bar{c}}{r}\right)'\right)
-\frac{2\alpha }{Y}\left\{\lambda\left(e'-\rho'_0
\right.\right.\right.\right.\\\nonumber &&\left.\left.\left.\left.
-\frac{2\alpha R'_0}{Y}\right)
+\lambda_1\left(e'p_{r0}
+e[p'_{r0}+\rho_0\frac{A'_0}{A_0}+p_{r0}(\frac{A'_0}{A_0}+\frac{2}{r}-\frac{2\alpha R'_0}{Y})]\right)\right\}
-\left(\lambda\right.\right.\right.\\\nonumber &&\left.\left.\left.
+\lambda_1\Gamma\frac{p_{r0}}{\rho_0+p_{r0}}\right)
\left\{
\rho_0\frac{2e}{Y}+\lambda_1\frac{\bar{c}}{r}(3\rho_0+p_{r0}+4p_{\perp0})+Y Z_{1p}\right\}_{,1}+\left\{\frac{A'_0}{A_0}\right.\right.\right.\\\nonumber
&&\left.\left.\left.
+\frac{2}{r}\frac{\lambda}{\lambda_1}+\Gamma\frac{p_{r0}}{\rho_0
+p_{r0}}(\frac{A'_0}{A_0}+\frac{2}{r}-\frac{2\alpha R'_0}{Y})
+\lambda_1\left(\Gamma\frac{p_{r0}}{\rho_0+p_{r0}}\right)'\right\}\left\{\frac{2e\rho_0}{Y}
\right.\right.\right.\\\label{B10} &&\left.\left.\left.
+\lambda_1\frac{\bar{c}}{r}(3\rho_0+p_{r0}+4p_{\perp0})
+(Y)Z_{1p}\right\}+\frac{2}{r}\frac{1}{\lambda_1}Z_3\right\}
+Z_{2p}\right]=0.
\end{eqnarray}

Matching conditions at boundary surface together with perturbed form of Eq.(\ref{f4}) can be
written in the simplified form as follows
\begin{equation}\label{66}
\ddot{D}(t)-Z_4(r) D(t)=0,
\end{equation}
provided that
\begin{equation}\label{67}
Z_4=\frac{rA_0^2}{Y\bar{c}-2\alpha
er}\left[\frac{2\alpha e}{Y}p_{\perp0}+\lambda\frac{\bar{c}}{r}(3\rho_0
+p_{r0}+4p_{\perp0})
+YZ_{1p}+\frac{Z_3}{\lambda_1}\right].
\end{equation}
The valid solution of Eq.(\ref{66}) turns out to be
\begin{equation}\label{68}
D(t)=-e^{\sqrt{Z_4}t}.
\end{equation}
The terms of $Z_4$ must be constrained in a way that all terms maintain
positivity. Impact of shear-free condition on dynamical instability of N and
pN regimes is covered in following subsections.

\subsection{Newtonian Regime}

In order to establish instability range in Newtonian era, we set $\rho_0\gg
p_{r0}$, $\rho_0\gg p_{\perp0}$ and $A_0=1,~B_0=1$. Insertion of these
assumptions and Eq.(\ref{68}) in Eq.(\ref{B10}) leads to the instability
condition, relating usual matter and dark source contribution as under
\begin{equation}\label{N}
\Gamma<\frac{Z_4X_3+X_4+\lambda\rho_0(X_2+YZ_{1p(N)})_{,1}+X_1 X_2-
\frac{2}{r\lambda_1}Z_{3(N)}+YZ_{2p(N)}}{\lambda_1p_{r0}X'_2
+\left\{p_{r0}\left(\frac{2\alpha R'_0}{Y}-\frac{2}{r}\right)\right\}X_2},
\end{equation}
where
\begin{eqnarray}\nonumber&&
X_1=(\lambda\rho'_0+\frac{2\lambda}{r\lambda_1}), \quad X_2 = \frac{2e}{Y}+3\lambda_1b,\quad \quad X_3 = -2\alpha^2b R'_0+Yb,
\\\nonumber && Z_4= \lambda_1\left[\rho_0 a'+2(p_{r0}+p_{\perp0})b'\right]+
\frac{2\alpha}{Y}\left[\lambda\left(\frac{2\alpha R'_0}{Y}-\rho'_0+e'\right)
\right.\\\nonumber &&\left.+\lambda_1\left\{p_{r0}+e[p'_{r0}+p_{r0}\left(\frac{2}{r}-\frac{2\alpha R'_0}{Y}\right)]\right\}\right].
\end{eqnarray}
The quantities $Z_{1p(N)}$ and $Z_{2p(N)}$ are terms of $Z_{1p}$ and $Z_{2p}$ belonging
to Newtonian era.
The gravitating source remains
stable in Newtonian approximation until the inequality for $\Gamma$ satisfies,
for which following constraints must be accomplished.
\begin{equation}\nonumber
 \quad 2\alpha R'_0<Y,
\quad \frac{2\alpha R'_0}{Y}>\rho_0'-e'
\end{equation}
The case when $\alpha\rightarrow0$ and $\lambda\rightarrow0$ leads to GR corrections
and results for $f(R)$ can be retrieved by setting $\lambda\rightarrow0$.

\subsection{Post Newtonian Regime}

We assume $A_0=1-\frac{m_0}{r}$ and $B_0=1+\frac{m_0}{r}$ to evaluate
stability condition in pN regime. On substitution of these assumptions in Eq.
(\ref{B10}), we have following inequality for $\Gamma$ to be fulfilled for
stability range.
\begin{equation}\label{zw}
\Gamma<\frac{Z_4X_5+X_6+\lambda\rho_0(X_7+YZ_{1p(PN)})_{,1}+X_8 X_7-
\frac{2}{r\lambda_1}Z_{3(PN)}+YZ_{2p(PN)}}{\lambda_1p_{r0}X'_7
+\left\{p_{r0}\left(\frac{m_0}{r(r-m_0)}
+\frac{2\alpha R'_0}{Y}+\frac{2}{r}\right)\right\}X_7},
\end{equation}
where
\begin{eqnarray}\nonumber
X_5&=&\frac{2\alpha r^2}{(r-m_0)^2}\left\{e'-\frac{r}{r+m_0}\left(bR'_0+2e\frac{m_0}{r}\right)\right\}+
Y\left[\frac{ r^2}{(r-m_0)^2}\left\{2\alpha e
\right.\right.\\\nonumber&&\left.\left.-Y\frac{\bar{c}}{r}\right\}-\frac{2\alpha(r+m_0)^2}{r^2}
\left\{\frac{er^2}{Y(r+m_0)^2}\right\}'\right],
\\\nonumber
X_6&=&\lambda_1\left\{\rho_0\left(\frac{ar}{r-m_0}\right)'-2(p_{r0}+p_{\perp0})b'\right\}
-\frac{2\alpha}{Y}\left[(\lambda_1 p_{r0}+\lambda)e'\right.\\\nonumber&&\left.
+e\left\{\lambda_1( p'_{r0}+\frac{\rho_0m_0}{r(r-m_0)})+p_{r0}\left(\frac{2}{r}-\frac{2\alpha R'_0}{Y}\right)\right\}
-\lambda\left(\rho'_0-\frac{2\alpha R'_0}{Y}\right)\right],\\\nonumber
X_7&=&\frac{2e}{Y}+\lambda_1b\left(2+\frac{r}{r+m_0}\right), \quad
X_8=\left(\frac{m_0}{r(r-m_0)}
+\frac{2\alpha R'_0}{Y}+\frac{2\lambda}{\lambda_1r}+\lambda\rho'_0\right).
\end{eqnarray}
$Z_{1p(PN)}$ and $Z_{2p(PN)}$ are terms of $Z_{1p}$ and $Z_{2p}$ that lie in
post-Newtonian era. The above inequality (\ref{zw}) holds for positive
definite terms and describe the stability range of subsequent evolution. The
positivity of each term appearing in (\ref{zw}) leads to following
restrictions
\begin{eqnarray}\nonumber&&
\frac{r}{r+m_0}(bR'_0+\frac{2em_0}{r})<e',\quad 2\alpha
e-Yb>\frac{(r^2-m_0^2)^2}{r^4}\left\{\frac{er^2}{Y(r+m_0)^2}\right\}'\\\nonumber &&
\left(\frac{ar}{r-m_0}\right)'>2(p_{r0}+p_{\perp0})b', \quad \rho'_0<\frac{2\alpha R'_0}{Y}.
\end{eqnarray}

\section{Concluding Remarks}

In this manuscript, we carried out a study on implications of shear-free
condition on stability of spherically symmetric anisotropic stars in $f(R,
T)$. Our exploration regarding viability of the $f(R,T)$ model reveals that
the selection of $f(R,T)$ model for dynamical analysis is constrained to the
form $f(R,T)=f(R)+\lambda T$, where $\lambda$ is arbitrary positive constant.
The restriction on $f(R,T)$ form originates from the complexities of
non-linear terms of trace in analytical formulation of field equations. The
model under consideration is of the form $f(R, T)=R+\alpha R^2+\lambda T$,
representing a viable substitute to dark source and the exotic matter
satisfying both viability criterion (positivity of radial derivatives upto
second order).

In $f(R,T)$ gravity, the non-minimal matter geometry coupling include the
terms of trace $T$ in action (\ref{1}) that is beneficent in the description
of quantum effects or so-called exotic matter. The components of modified
field equations together with the implementation of shear-free condition are
developed in section \textbf{2}. Further conservation laws are considered to
arrive at dynamical equations by means of Bianchi identities. These equations
are utilized to estimate the variations in gravitating system with the
passage of time.

The complexities of more generic analytical field equations are dealt by
using linear perturbation of physical quantities. Perturbation scheme induce
significant ease in the description of dynamical system, rather to present
stability analysis by means of numerical simulations. The analytic approach
we have employed here is more general and substantially important in
explorations regarding structure formation. The perturbed shear-free
condition together with the dynamical and field equations lead to the
evolution equation, relating $\Gamma$ with the usual and dark source terms.
It is found that the induction of trace of energy momentum tenor in action
(\ref{1}) contributes positive addition to $\Gamma$, that slows down
subsequent evolution considerably.

The outcome of gravitational evolution is size dependent as well as other
physical aspects such as isotopy, anisotropy, shear, radiation, dissipation
etc. The instability range for N and pN approximations is considered that
impose some restrictions on the physical variables. It is observed that the
terms appearing in $\Gamma$ are less constrained for both the regimes (N and
pN) in comparison to the anisotropic sources \cite{1}. Thus, shear-free
condition benefits in more stable anisotropic configurations. Corrections to
GR and $f(R)$ establishments can be made by setting $\alpha\rightarrow0$,
$\lambda\rightarrow0$ and $\lambda\rightarrow0$ respectively. The local
isotropy of the model can be settled by assuming $p_r=p_\perp=p$. The
extension of this work for shearing expansion free evolution of anisotropic
spherical and cylindrical sources is in process.

\section{Appendix}

\begin{eqnarray}\setcounter{equation}{1}\nonumber
Z_1(r,t)&=&f_R A^2\left[\left\{\frac{1}{f_R A^2}\left(\frac{f-Rf_R}{2}
-\frac{3\dot{f_R}}{A^2}\frac{\dot{B}}{B}-\frac{f'_R}{B^2}\left(\frac{B'}{B}
-\frac{2C'}{C}\right)+\frac{f''_R}{B^2}\right)\right\}_{,0}\right.\\\nonumber &&\left.
+\left\{\frac{1}{f_RA^2B^2}\left(\dot{f'_R}-\frac{A'}{A}\dot{f_R}
-\frac{\dot{B}}{B}f_R'\right)\right\}_{,1}\right]-\frac{\dot{f_R}}{A^2}\left\{
\left(3\frac{\dot{B}}{B}\right)^2+\frac{9\dot{A}}{A}\frac{\dot{B}}{B}
\right\}
\\\nonumber&&
+\frac{3\ddot{f_R}}{A^2}\frac{\dot{B}}{B}+\frac{\dot{A}}{A}(f-Rf_R)
-\frac{2f'_R}{B^2}\left\{\frac{\dot{A}}{A}\left(\frac{B'}{B}
-\frac{C'}{C}\right)+\frac{\dot{B}}{B}\left(\frac{3A'}{A}-\frac{2C'}{C}\right.\right.\\\nonumber
&&\left.\left.
+\frac{B'}{B}\right)\right\}
+\frac{f_R''}{B^2}\left(\frac{2\dot{A}}{A}+\frac{\dot{B}}{B}\right)+\frac{1}{B^2}\left(\dot{f'_R}-\frac{A'}{A}\dot{f_R}
\right)\left(\frac{3A'}{A}
+\frac{B'}{B}+\frac{2C'}{C}\right),\\\label{B3}
&&
\\\nonumber
Z_2(r,t)&=&f_RB^2\left[\left\{\frac{1}{f_RA^2B^2}\left(\dot{f_R}'-\frac{A'}{A}\dot{f_R}
-\frac{\dot{B}}{B}f_R'\right)\right\}_{,0}+\left\{\frac{1}{f_RB^2}\left(\frac{Rf_R-f}{2}
\right.\right.\right.
\\\nonumber &&\left.\left.\left.-\frac{\dot{f_R}}{A^2}\left(\frac{\dot{A}}{A}-\frac{2\dot{C}}{C}\right)-\frac{f'_R}{B^2}\left(\frac{A'}{A}
+\frac{2C'}{C}\right)+\frac{\ddot{f_R}}{A^2}\right)\right\}_{,1}
\right]+(Rf_R-f)\frac{B'}{B}
\\\nonumber
&&-\frac{1}{A^2}\frac{\dot{f_R}}{A^2}\left\{\frac{A'}{A}\left(\frac{\dot{A}}{A}+\frac{\dot{B}}{B}\right)
+\frac{B'}{B}\left(\frac{\dot{A}}{A}-\frac{2\dot{C}}{C}\right)\right\}
+\left(\frac{\dot{A}}{A}+\frac{5\dot{B}}{B}\right)\left(\dot{f_R}'\right.\\\nonumber
&&\left.-\frac{A'}{A}\dot{f_R}
-\frac{\dot{B}}{B}f_R'\right)-\frac{f'_R}{B^2}\left\{\frac{A'}{A}\left(\frac{A'}{A}+\frac{3B'}{B}\right)
+\frac{2C'}{C}\left(\frac{3B'}{B}
+\frac{C'}{C}\right)\right\}
\\\label{B4} &&
+\frac{\ddot{f_R}}{A^2}\left(\frac{A'}{A}+\frac{2B'}{B}\right)
+\frac{f''_R}{B^2}\left(\frac{A'}{A}
+\frac{2C'}{C}\right).
\end{eqnarray}
\begin{eqnarray}\nonumber
Z_{1p}&=& 2\alpha A_0^2\left[\frac{1}{A_0^2B_0^2Y}\left\{e'-e\frac{A'_0}{A_0}-\frac{b}{B_0}R'_0\right\}\right]_{,1}
+\frac{1}{Y}\left[e-[\lambda T_0
\right.\\\nonumber && \left.-\alpha R_0^2]\left(\frac{a}{A_0}+\frac{e}{Y}\right)
+\frac{2\alpha}{B_0^2}\left\{\left(\frac{B'_0}{B_0}-\frac{2}{r}\right)\left(e'-2R'_0\left(\frac{a}{A_0}
+\frac{b}{B_0}\right)\right.\right.\right.\\\nonumber & &\left.\left.\left.
+\frac{2 \alpha e}{Y}R'_0\right)
+ R''_0 \left(\frac{2a}{A_0}+\frac{b}{B_0}\right)-2R'_0\left(\frac{b}{B_0}\left(\frac{2A'_0}{A_0}+
\frac{B'_0}{B_0}+\frac{1}{r}\right)\right.\right.\right.\\\label{Z1p} & &\left.\left.\left.+\frac{\bar{c}}{r}\left(\frac{A'_0}{A_0}-\frac{3}{r}\right)\right)
+\left(e'-e\frac{A'_0}{A_0}\right)\left(\frac{3A'_0}{A_0}+
\frac{B'_0}{B_0}+\frac{2}{r}\right)\right\}\right]
\\\nonumber
Z_{2p}&=&B_0^2Y\left[\frac{1}{B_0^2Y}\left\{e-\frac{2\alpha}{B_0^2}\left\{
\left(\frac{A'_0}{A_0}+\frac{2}{r}\right)\left(e-[\frac{2 \alpha e}{Y}+\frac{4b}{B_0}]R'_0\right)
+R'_0[\left(\frac{a}{A_0}\right)'
\right.\right.\right.\\\nonumber & &\left.\left.\left.
+\left(\frac{\bar{c}}{r}\right)']\right\}-[\lambda T_0
-\alpha R_0^2]\left(\frac{b}{B_0}+\frac{e}{Y}\right)\right\}\right]_{,1}
+bB_0Y\left[\frac{1}{B_0^2Y}\left\{\lambda T_0
-\alpha R_0^2
\right.\right.\\\nonumber &&\left.\left.
-\frac{4\alpha}{B_0^2}\left(\frac{A'_0}{A_0}+\frac{2}{r}\right)R'_0\right\}\right]_{,1}
+\frac{2\alpha}{B_0^2}\left[R''_0\left\{\left(\frac{a}{A_0}\right)'-2\left(\frac{A'_0}{A_0}
+\frac{2}{r}\right)\left(\frac{b}{B_0}+\frac{e}{Y}\right)
\right.\right.\\\nonumber &&\left.\left.+\left(\frac{\bar{c}}{r}\right)'
\right\}-R'_0\left\{\frac{A'_0}{A_0}\left[2\left(\frac{a}{A_0}\right)'+3\left(\frac{b}{B_0}\right)'\right]
+3\frac{B'_0}{B_0}\left[\left(\frac{a}{A_0}\right)'+2\left(\frac{\bar{c}}{r}\right)'\right]
\right.\right.\\\nonumber & &\left.\left.+
\frac{2}{r}\left[\left(3\frac{b}{B_0}\right)'+2\left(\frac{\bar{c}}{r}\right)'\right]\right\}+\left(\frac{2b}{B_0}R'_0
-e\right)
\left\{3\frac{B'_0}{B_0}\left(\frac{A'_0}{A_0}+\frac{2}{r}\right)
+\left(\frac{A'_0}{A_0}\right)^2
\right.\right.\\\label{Z2p} & &\left.\left.+\frac{2}{r^2}\right\}\right]+e\frac{B'_0}{B_0}-[\lambda T_0
-\alpha R_0^2]\left(\frac{b}{B_0}+\frac{2e}{Y}\frac{B'_0}{B_0}\right)
\end{eqnarray}
\begin{eqnarray}\nonumber
Z_{3}&=&\frac{Y}{B_0^2}\left[\frac{a''}{A_0}+\frac{\bar{c}''}{r}-\frac{A''_0}{A_0}
\left(\frac{a}{A_0}+\frac{2b}{B_0}\right)+\frac{A'_0}{A_0}\left\{\frac{2b}{B_0}\left(\frac{B'_0}{B_0}
-\frac{1}{r}\right)+\left(\frac{\bar{c}}{r}\right)'\right.\right.\\\nonumber & &\left.\left.
-\left(\frac{b}{B_0}\right)'\right\}
+\frac{B'_0}{B_0}\left\{\frac{2bB_0'}{rB_0}-\left(\frac{a}{A_0}\right)'-\left(\frac{\bar{c}}{r}\right)'\right\}+
\frac{1}{r}\left\{\left(\frac{a}{A_0}\right)'-\left(\frac{b}{B_0}\right)'\right\}
\right]\\\nonumber & &
-\frac{2\alpha e}{Y}\left\{\frac{\lambda T_0
-\alpha R_0^2}{2}-\frac{2\alpha}{B_0^2}\left(R'_0\left(\frac{A'_0}{A_0}-\frac{B'_0}{B_0}
+\frac{1}{r}\right)-R''_0\right)\right\}-\frac{2\alpha}{B_0^2}\left\{e''\right.\\\label{Z3} & &\left.
+\frac{2b}{B_0}R''_0+\left(\frac{A'_0}{A_0}-\frac{B'_0}{B_0}
+\frac{1}{r}\right)\left(\frac{2b}{B_0}R'_0-e'\right)\right\}
\end{eqnarray}

\vspace{.25cm}

{\bf Acknowledgment}

\vspace{.25cm}

The authors thank the anonymous referee for fruitful comments and
suggestions.

\end{document}